# EMTk – The Emotion Mining Toolkit

Fabio Calefato, Filippo Lanubile, Nicole Novielli, Luigi Quaranta
University of Bari Aldo Moro
Bari, Italy
{fabio.calefato | filippo.lanubile | nicole.novielli}@uniba.it, l.quaranta6@studenti.uniba.it

***Abstract*—The Emotion Mining Toolkit (EMTk) is a suite of modules and datasets offering a comprehensive solution for mining sentiment and emotions from technical text contributed by developers on communication channels. The toolkit is written in Java, Python, and R, and is released under the MIT open source license. In this paper, we describe its architecture and the benchmark against the previous, standalone versions of our sentiment analysis tools. Results show large improvements in terms of speed.**

***Keywords—sentiment analysis, emotion mining, social software engineering***

## I. Introduction

Sentiment analysis has become a crucial task in a technical domain such as software engineering (SE). To date, several off-the-shelf tools are available for classifying the sentiment polarity of text, that is the positive, negative, or neutral semantic orientation, or recognizing specific emotions, such as joy, love, and anger. In this paper, we present the Emotion Mining Toolkit (EMTk),[1] a comprehensive, SE-specific solution for mining polarity and emotions from technical text contributed by developers on communication channels (e.g., issue comments, Q&A sites).

EMTk integrates two tools, namely EmoTxt [1] and Senti4SD [2], which have been developed during 2016 and released as standalone applications in mid-2017 under the MIT license.

EmoTxt [1] supports the recognition in technical text of six specific emotions (i.e., joy, love, surprise, anger, sadness, and fear), according to the framework by Shaver et al. [9]. EmoTxt leverages a suite of features capturing the presence of lexical cues conveying emotions in the input text:

1. *Uni- and bi-grams*, modeled using a tf-idf weighting schema.
2. *Emotion lexicon*, captured by using WordNet Affect [10].
3. *Politeness*, measured by the tool of Danescu et al. [3].
4. *Mood*, measured by the tool of De Smedt et al. [4].

Senti4SD [2] is a polarity classifier trained to support sentiment analysis in developers' communication channels using supervised machine learning. It leverages a suite of features based on Bag of Words (BoW), sentiment lexicons, and semantic features based on word embedding. Compared to the performance obtained on the same Stack Overflow test set by SentiStrength [11], a general-purpose sentiment analysis tool that has been widely adopted before the development and release of SE-specific classifiers, Senti4SD reduces the misclassifications of neutral and positive posts as emotionally negative. Senti4SD achieves good performance ($F_1$=.84) with a minimal set of ~400 training documents. Both EMTk and Senti4SD provide a training method that enables the customization of the classifier using a gold standard as input.

Compared to the usage of the two separate standalone tools, EMTk is easy to install, because it is distributed as a Docker container, and faster, because it leverages parallel computation.

In the rest of this paper, we first describe the architecture of the toolkit with its main modules and experimental datasets (Sect. II). Then, we show how to install and run EMTk (Sect. III). Also, we benchmark the toolkit against the standalone versions of EmoTxt and Senti4SD to show how EMTk increases the speed (Sect. IV). Finally, we draw conclusions (Sect. V).

## II. Toolkit Architecture

EMTk consists of two modules, one for the mining of polarity and one for emotions. These two modules are based on our previously developed tools, respectively Senti4SD and EmoTxt, whose code has been refactored and optimized. Below, we briefly report their structure and the performance optimization.

### A. Emotion Classification Module

To develop the emotion module of EMTk, we built upon the code base of EmoTxt, which was written in Java 8, Python 2.7, and R. First, we reduced the technical debt by performing mostly structural refactoring. Then, based on consistent feedback about the slowness of the tool, we sought possible bottlenecks and found that the Stanford CoreNLP library [6] was one, due to the inherently time-consuming task of parsing. Specifically, the output of the parsing module is required in input by the *politeness* [3] and *mood* modules [4] in EmoTxt. As such, during both training and classification in EMTk, we made optional the politeness and mood features. When they are disabled, EMTk uses the Stanford CoreNLP to perform only tokenization and sentence split. Still, the largest performance improvement was obtained by leveraging the parallel streams API available in Java 8. Further information regarding the gain in performance and its measures is reported in Sect. IV. Finally, regarding the Java part of the module, we have introduced the use of Apache Maven[2] for managing the building and the dependencies from external libraries.

[1] https://collab-uniba.github.io/EMTk

[2] https://maven.apache.org

TABLE I. BREAKDOWN OF THE EXPERIMENTAL DATASETS PROVIDED WITH THE TOOLKIT WITH RESPECT TO EMOTIONS.

| Dataset | Love | Joy | Surprise | Anger | Fear | Sadness |
|---|---|---|---|---|---|---|
| Stack Overflow (N=~4,800) | 1,220 (25%) | 491 (10%) | 45 (1%) | 882 (18%) | 230 (5%) | 106 (2%) |
| Jira (N=~4,000) | 166 (4%) | 124 (3%) | NA | 324 (8%) | NA | 302 (7%) |

### B. *Polarity Classification Module*

To develop the polarity module of EMTk, we built upon the codebase of Senti4SD, which is developed for the most part in Java 8, with a few components written in R. The main complaint of the Senti4SD adopters was about slowness and then we opted to leverage parallel computation to increase its speed.

We first refactored the code to reduce the technical debt; then, we used Apache Maven for better dependency and build management. Finally, as all the computation in Senti4SD was performed sequentially, we redesigned the optimized core to implement the actor system in Akka [5], a toolkit for building highly-concurrent, message-driven applications for Java. An *actor system* (or *actor model*) is a hierarchy of actors, that are objects which encapsulate state and behavior, and communicate exclusively by exchanging messages placed into the mailbox of the recipient actor. In the new redesigned module (see Fig. *1*), we have a *master* actor at the root of the hierarchy, which starts and process and oversees four types of actors, namely the *reader* and the *writer*, which handle the respective I/O operations, and the *router*, which dispatches messages (i.e., operation descriptions) to the *worker* actors. *Workers* handle the core of the computation and forward the results as messages to the *writer* actor so that they can be stored. In our system, actors are implemented as threads. After experimenting with a growing number of actors, we found that the optimal performance is achieved with 8 threads (i.e., matching the number of the CPU cores available on the host). Further information regarding the gain in performance and its measures is reported in Sect. IV.

### C. *Experimental Datasets*

Along with our toolkit, we also package two experimental datasets, one from Stack Overflow and one from Jira. Once the Docker container is installed (see Sect. III), both experimental datasets are available at `/datasets/`.

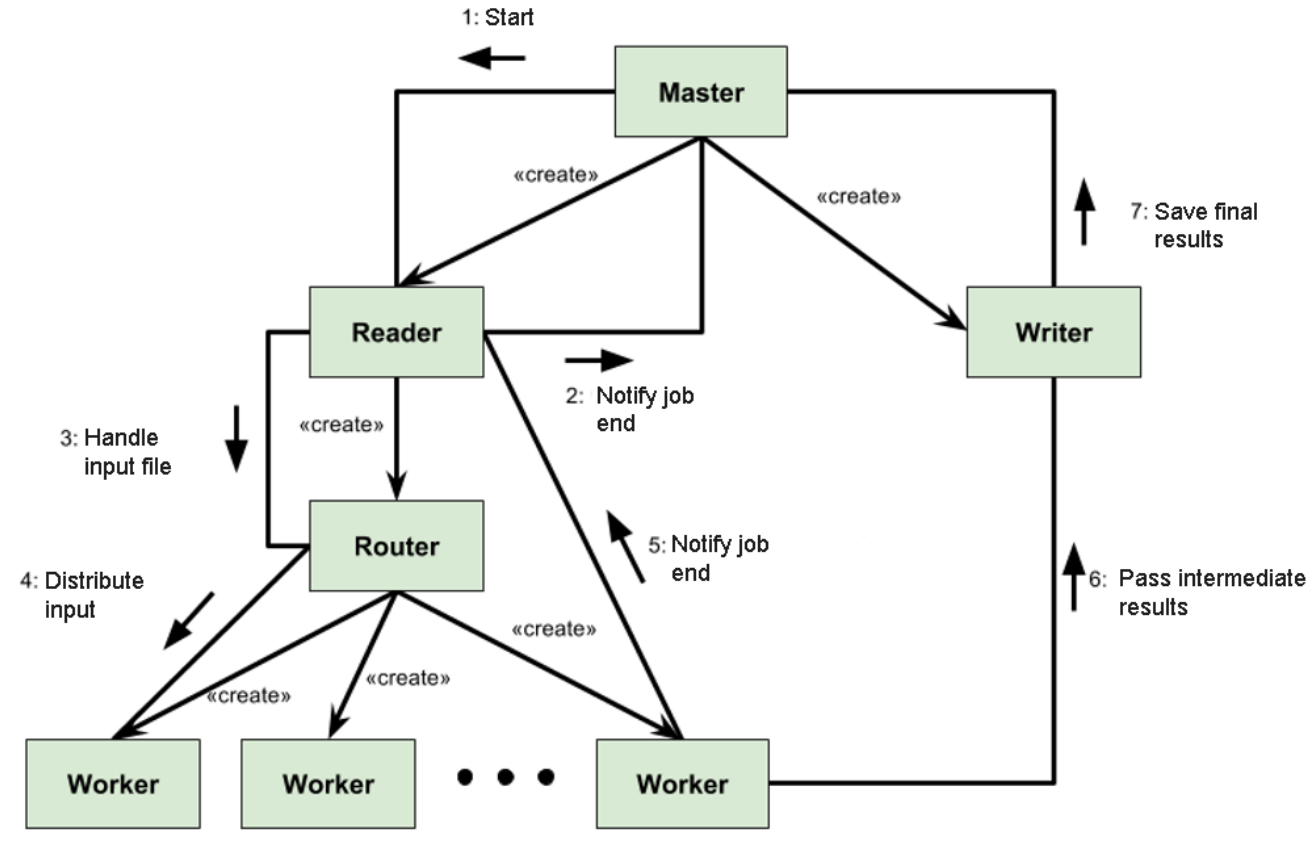

Fig. 1- Hierarchy of the actor system implemented in the polarity module of EMTk.

The Stack Overflow dataset [7] consists of 4,800 posts, in the form of questions, answers, and comments, manually annotated by twelve coders with a background in Computer Science. The coders were trained to explicitly indicate a label for each post according to the emotion detected, based on the six basic emotions framework (i.e., love, joy, surprise, anger, fear, and sadness) proposed by Shaver et al. [9]. Each post was annotated by three raters and, for all posts in the dataset, both the set of individual annotations provided by the raters (i.e., the indication of presence or absence for each emotion) and the gold label obtained by applying majority voting are made available.

The Jira dataset is provided by Ortu et al. [8]. It includes ~4,000 issue comments and sentences authored by software developers of popular open source software projects, such as Apache and Spring. The Jira dataset was annotated by three raters based on the same emotion framework by Shaver et al. [9]. However, the comments were manually labeled with only 4 emotions, namely love, joy, anger, and sadness.

In TABLE I, we report the breakdown of the distribution of the gold labels for each emotion in the two datasets. Note that the emotion labels are not mutually exclusive, that is, text may have multiple labels when found to convey more than one emotion. To make the two datasets annotated with the emotion labels also compatible with the EMTK module for analyzing polarity (i.e., the positive, negative, or neutral valence conveyed by texts), we translated the positive emotions (i.e., love and joy) into a positive polarity label. Similarly, sadness, anger, and fear were mapped to the negative polarity class. Instead, surprise cases were discarded as this emotion label can be either considered positive or negative, depending on the expectations of the author of a text. Finally, the absence of emotions defined neutral cases. The Stack Overflow gold standard resulting from this procedure is well-balanced (see TABLE II), with 35% of posts conveying positive emotions, 27% presenting negative emotions, and 38% of posts labeled as neutral, denoting absence of emotions. Instead, the Jira gold standard is not as well-balanced, with 19% of posts conveying positive emotions, 13% conveying negative emotions and 68% labeled as neutral.

## III. INSTALLATION AND EXECUTION

ETMk code is available from GitHub. However, to ease the installation and setup costs, we have packaged it as a lightweight

TABLE II. BREAKDOWN OF THE EXPERIMENTAL DATASETS PROVIDED WITH THE TOOLKIT WITH RESPECT TO POLARITY.

| Dataset | Positive | Neutral | Negative |
|---|---|---|---|
| Stack Overflow | 35% | 38% | 27% |
| Jira | 19% | 68% | 13% |

Docker container and published on Docker-Hub[3]. As such, anyone interested in using EMTk must first install Docker;[4] then, to install the latest version of the EMTk container image, run the following from the command line:

```
$ docker pull collabuniba/emtk
$ docker run --rm -v <sharedFolderPath>:/shared -ti collabuniba/emtk
```

where <sharedFolderPath> is the path to the folder on the host that will be shared with the container to allow file exchange at runtime.

The second line logs you in the container's shell environment (>). From there, to execute the polarity module, run:

```
> emtk polarity -F A -i input.csv -oc output.csv -vd 600
[-W dsm.bin] [-L] [-ul unigramList -bl bigramList]
```

where:

- -F {A, S, L, K}, feature to evaluate A for All, S for Semantic, L for Lexicon, K for Keyword.
- -i <input.csv>: the input data to classify.
- -oc <output.csv>: the resulting predictions.
- -vd <N>: the vector size.
- -W <dsm.bin>: [optional] the wordspace to use.
- -L: [optional] if present, the input corpus comes with a gold label in the label column.
- -ul <filename>: [optional] the unigram's list.
- -bl <filename>: [optional] the bigram's list.

Users can test-drive the polarity module by using the file /polarity/Sample.csv, containing only a handful of documents.

Regarding the emotion classifier module, in the following, we show first how to train a new model and, then, how to test it on unseen data. To train a new model on a training set, run:

```
> emtk emotions train -i file.csv –p -d delimiter [-g] -e emotion
```

where:

- -i <file.csv>: the corpus to be classified, encoded in UTF-8 without BOM and with the following format:
```
id;label;text
…
22;NO;"""Excellent! This is exactly what I needed.
  Thanks!"""
23;YES;"""FEAR!!!!!!!!!!!"""
…
```
- -p: enables the extraction of features regarding politeness, mood and modality.
- -d {c, sc}: the delimiter used in the csv file, where c stands for comma and sc for semicolon.
- -e {joy, anger, sadness, love, surprise, fear}: the emotion to be detected.

As a result, the script will generate an output folder at the location /polarity/training_<file.csv>_<emotion>/, containing:

- n-grams/: a subfolder containing the extracted n-grams.
- idfs/: a subfolder containing the IDFs computed for n-grams and WordNet Affect emotion words.
- feature-<emotion>.csv: a file with the features extracted from the input corpus and used for training the model.
- liblinear/DownSampling/ and liblinear/NoDownSampling/, two folders each containing:
    - trainingSet.csv and testSet.csv.
    - eight models trained with liblinear model_<emotion>_<ID>.Rda, where ID refers to the liblinear model (with values in {0, ..., 7}).
    - performance_<emotion>_<IDMODEL>.txt, a file containing the results of the parameter tuning for the model (cost), the confusion matrix, and the Precision, Recall, and F-measure for the best cost for the specific <emotion>.
    - predictions_<emotion>_<IDMODEL>.csv, containing the test instances with the predicted labels for the specific <emotion>.

Finally, to execute the classification task, run:

```
> emtk emotions classify -i file.csv -p -d delimiter -e
emotion [-m model] [-f /path/to/.../idfs] [-o
/path/to/.../ngrams] [-l]
```

where:

- -i <file.csv>: same as above.
- -p: enables the extraction of features regarding politeness, mood and modality.
- -d {c, sc}: same as above.
- -e {joy, anger, sadness, love, surprise, fear}: same as above.
- -m model: [optional] the model file learned during the training step; if not specified, as default the model learned on the Stack Overflow gold standard will be used.
- -f /path/to/.../idfs: [optional] with custom models, also the path to the folder containing the dictionaries with IDFs computed during the training step is required; the folder must include IDFs for n-grams (uni- and bi-grams) and the WordNet Affect lists of emotion words.
- -o /path/to/.../ngrams: [optional] with custom models, also the path to the folder containing the dictionaries extracted during the training step; the folder must include n-grams (i.e., UnigramsList.txt and BigramsList.txt).
- -l: [optional] if present, the input corpus comes with a gold label in the column label.

As a result, the script will create an output folder at the location /emotions/classification_<file.csv>_<emotion>, containing:

- predictions_<emotion>.csv: a csv file, containing a binary prediction (yes/no) for each line of the input corpus:
```
id;predicted
…
22;NO
23;YES
…
```
- performance_<emotion>.txt: a file containing several performance metrics (Precision, Recall, F1, confusion matrix), created only if the input corpus <file.csv> contains the column label.

[3] https://hub.docker.com/r/collabuniba/emtk

[4] www.docker.com

TABLE III. RESULTS OF THE BENCHMARK BETWEEN EMTK (WITH 8 CONCURRENT THREADS) AND THE SEQUENTIAL STANDALONE TOOLS.

| Benchmark | Polarity | | Emotion | |
|---|---|---|---|---|
| | ***Senti4SD*** | ***EMTk polarity*** | ***EmoTxt*** | ***EMTk emotion*** |
| **Execution time*** | 56m 46s | 1m 20s | 1h 4m 41s | 33m 3s ° |
| | | | | 18m 53s § |
| **Speedup**** | 42.58 | | 1.96 ° | |
| | | | 3.43 § | |

* the smaller, the better
** the larger, the better
° w/ politeness and mood features
§ w/o politeness and mood features

Users can test-drive the emotion classification module by using the file `/emotions/sample.csv`, containing only a handful of documents. Other, more complex sample datasets are available at `/emotions/java/DatasetSO/StackOverflowCSV`.

To use the EMTk modules with custom datasets, users must access the folder `/shared/`, which is mounted specifying the `-v` option in the `docker run` command and can be used to share input and output files. For further instructions, users can refer to the EMTk documentation available on the website.

## IV. BENCHMARK

Data in TABLE III refer to the execution time as measured on an x86_64 virtual machine provided by the data center of the University of Bari, with 8 Intel i7-2600K CPUs clocked at 2 GHz and 32 gigabytes of RAM. Because the test environment is equipped with 8 CPUs, in our tests we achieved the best results in a setting with as many concurrent threads. This is because, with fewer threads, the CPUs were underused; instead, with more than 8 threads, performance decreases due to I/O bottlenecks. For the sake of space, here we report only the metrics collected in the best configuration.

### A. Metrics

To measure the relative performance of our toolkit with parallel computation against the sequential execution, we use two metrics, execution time and speedup. These metrics are intended to assess how the new EMTk modules compare against the previous versions (i.e., the standalone tools). While the execution time metric is intuitive, speedup is defined as $S = \frac{T_1}{T_P}$, where $P$ is the number of processors, $T_1$ is the execution time in seconds of the previous version, with sequential configuration (i.e., with 1 processor), and $T_P$ is the execution time of the new parallel configuration with $P$ processors.

### B. Results

Regarding the polarity, the benchmarking task consisted in classifying the polarity from a sample of ~3,100 random documents selected from the Stack Overflow dataset. We observed a large decrease in the execution time between the new parallel version (1 min 20 sec) and the old sequential one (56 min 46 sec), which correspond to a speedup of S=42.58. We also repeated the experiment using the ~5,900 documents from the entire Jira dataset by Ortu et al., obtaining even better results, with a large decrease in execution time (2 min 24 sec vs. 1h 49 min 59 sec, S=6,540.41).

As for the emotion mining, the benchmarking task was to identify input texts expressing love in a sample dataset of 4,800 entries. An even larger decrease of execution time is observed in this case, as the standalone module would take over one hour (1h 4m 41 s), whereas the new EMTk module only takes a little more than half an hour (33m 3s), thus resulting in a speedup of 1.96. The new emotion module is even faster if the user wants to disable the features to compute the politeness and mood, which further halves the execution time (18m 53s, S=3.43).

## V. CONCLUSIONS AND FUTURE WORK

We presented EMTk, a comprehensive, SE-specific toolkit for emotion mining from technical text snippets. The EMTk modules are based on the refactoring of EmoTxt and Senti4SD tools, which had already been used by researchers. Compared to the standalone tools, the new redesigned and integrated version is faster and easier to install.

## ACKNOWLEDGMENTS

This work is partially funded by projects "Creative Cultural Collaboration" (C3) and "Electronic Shopping & Home delivery of Edible goods with Low environmental Footprint" (E-SHELF), under the Apulian INNONETWORK programme, Italy. The computational work has been executed on the IT resources made available by two projects, ReCaS and PRISMA, funded by MIUR under the program "PON R&C 2007-2013."